\documentclass[english,aip, jcp, reprint, onecolumn]{revtex4-1}
\usepackage[T1]{fontenc}
\usepackage[latin9]{inputenc}
\usepackage{multirow}
\usepackage{amsmath}
\usepackage{graphicx}

\makeatletter

\providecommand{\tabularnewline}{\\}

\usepackage{tikz}

\makeatother

\usepackage{babel}
\begin{document}

\title{Bridge functional for the molecular density functional theory with
consistent pressure and surface tension and its importance for solvation
in water}

\author{Cédric Gageat}

\affiliation{PASTEUR, Département de chimie, École normale supérieure, UPMC Univ. Paris
06, CNRS, PSL Research University, 75005 Paris, France}

\affiliation{Sorbonne Universités, UPMC Univ. Paris 06, École normale supérieure,
CNRS, PASTEUR, 75005 Paris, France}

\author{Luc Belloni}

\affiliation{LIONS, NIMBE, CEA, CNRS, Université Paris-Saclay, 91191 Gif-sur-Yvette,
France}

\author{Daniel Borgis}

\affiliation{PASTEUR, Département de chimie, École normale supérieure, UPMC Univ. Paris
06, CNRS, PSL Research University, 75005 Paris, France}

\affiliation{Sorbonne Universités, UPMC Univ. Paris 06, École normale supérieure,
CNRS, PASTEUR, 75005 Paris, France}

\affiliation{Maison de la Simulation, CEA, CNRS, Univ. Paris-Sud, UVSQ, Université
Paris-Saclay, 91191 Gif-sur-Yvette, France}

\author{Maximilien Levesque}

\affiliation{PASTEUR, Département de chimie, École normale supérieure, UPMC Univ. Paris
06, CNRS, PSL Research University, 75005 Paris, France}

\affiliation{Sorbonne Universités, UPMC Univ. Paris 06, École normale supérieure,
CNRS, PASTEUR, 75005 Paris, France}
\email{maximilien.levesque@ens.fr}

\selectlanguage{english}%
\begin{abstract}
We address the problem of predicting the solvation free energy and
equilibrium solvent density profile in fews minutes from the molecular
density functional theory beyond the usual hypernetted-chain approximation.
We introduce a bridge functional of a coarse-grained, weighted solvent
density. In few minutes at most, for solutes of sizes ranging from
small compounds to large proteins, we produce (i) an estimation of
the free energy of solvation within 1~kcal/mol of the experimental
data for the hydrophobic solutes presented here, and (ii) the solvent
distribution around the solute. Contrary to previous propositions,
this bridge functional is thermodynamically consistent in that it
produces the correct liquid-vapor coexistence and the experimental
surface tension. We show this consistency to be of crucial importance
for water at room temperature and pressure. This bridge functional
is designed to be simple, local, and thus numerically efficient. Finally,
we illustrate this new level of molecular theory of solutions with
the study of the hydration shell of a protein.
\end{abstract}
\maketitle
Molecules in solution, may they be small chemical compounds like drugs
or large biomolecules, behave differently from vacuum conditions.
How to treat the solvent, the embedding medium, in numerical simulations
is challenging. Since solvent molecules are the most abundant species
in the system, most of the simulation time is dedicated to computing
solvent-solvent interactions. In that sense, they are the limiting
constituant of the simulation. To circumvent this problem, implicit
models have been proposed, that mimick the solvent as a continuous
medium of macroscopic permittivity and local polarizability \cite{tomasi_molecular_1994,roux_implicit_1999,cramer1999implicit,cramer_universal_2008}.
Implicit models lack the molecular nature of the interactions, like
hydrogen bonding or steric effect and impede the rigorous evaluation
of entropic contributions. Nevertheless, good parameterizations allowed
their rapid and glorious development, sometimes with predicted free
energies in good agreement with simulations for a numerical cost reduced
to few seconds, orders of magnitude below simulations run-times. Liquid
state theories have been another approach, especially the integral
equation theory in the molecular picture \cite{blum_invariant_1972,blum_invariant_1972-1}
in the first place. Molecular integral equations, within the so-called
Molecular Ornstein-Zernike (MOZ) formalism, are nevertheless difficult,
sensible to numerical instabilities and extended only recently to
three dimensional systems \cite{Beglov-Roux96,beglov_integral_1997,IshizukaYoshida2013}.
They do not seem well suited to arbitrarily complex solutes in 3 dimensions,
even if it is still an especially dynamic area of research. An approximation
of this method was also developed, the reference interaction site
model (RISM)\cite{chandler-RISM,hirata_application_1982}, and then
derived and adapted to three-dimensional solutes with 3D-RISM \cite{beglov_integral_1997,hirata_molecular_2003,liu_site_2013}.
RISM and 3D-RISM are having large successes since they predict solvation
energies and profiles with better accuracy than usual continuous methods,
as shown recently for small neutral molecules~\cite{truchon_cavity_2014,Sheng2016,luchko2016},
large protein-ligand complexes~\cite{Imai2009} and even ions~\cite{palmer2016}.
Nevertheless, they rely on considering molecules as a set of sites
correlated together and are thus not diagramaticaly consistent. The
molecular density functional theory is yet another route. It is diagrammaticaly
consistent, has deep connections with the integral equation theories,
but is significantly less sensitive to numerical difficulties since
it relies on functional minimization, a confortable variational problem.
Still, finding the perfect functional is difficult, not to say impossible,
but nevertheless an ongoing project~\cite{ramirez_density_2005,gendre_classical_2009,zhao_molecular_2011,borgis_molecular_2012,levesque_scalar_2012,levesque_solvation_2012,jeanmairet_molecular_2013-1}
that is gaining momentum. Readers interested in the deep roots of
liquid state theories and classical density functional theory should
refer to \cite{evans_nature_1979,evans_density_2009,henderson_fundamentals_1992}. 

The theoretical work presented therein lies within the classical density
functional theory framework, more specifically the molecular density
functional theory (MDFT). MDFT consists in minimizing a free energy
functional of the solvent density. It is a cousin of the Khon-Sham
DFT for electrons \cite{kohn_self-consistent_1965}, expressed for
a classical molecular solvent density. For atomic fluids, this density,
$\rho(\boldsymbol{r})$, is expressed in number of solvent molecules
per volume unit. For molecular solvents like water, the molecular
density, $\rho\left(\boldsymbol{r},\omega\right)$, depends upon the
position and upon the molecular orientation. The notation $\omega$
refers to the triplet of Euler angles that are necessary and sufficient
to describe the orientation of any rigid body in three dimensions.
The free energy of solvation of an embedded molecule (the solute)
can be written as 
\begin{equation}
\Delta G_{\textrm{solv}}=\mathrm{min}\left(\mathcal{F}_{\mathrm{exc}}+\mathcal{F}_{\mathrm{ideal}}+\mathcal{F}_{\mathrm{ext}}\right),
\end{equation}
where $\mathcal{F}_{\mathrm{exc}}$, $\mathcal{F}_{\mathrm{ideal}}$
and $\mathcal{F}_{\mathrm{ext}}$ are the excess, ideal and external
functionals of $\rho\left(\boldsymbol{r},\omega\right)$, respectively.
The min is the variational minimum with respect to $\rho\left(\boldsymbol{r},\omega\right)$.
The whole MDFT procedure can be summarized as how to find the spatial
and angular density that minimizes the sum of these three contributions.
Knowing the functionals, the \textquotedbl{}minimizing\textquotedbl{}
density, by virtue of Kohn, Sham and Mermin's theorems, is also the
equilibrium density of the solvent around the solute. $\rho_{\textrm{equ.}}\left(\boldsymbol{r},\omega\right)$
and $\Delta G_{\textrm{solv}}$ are the two natural outputs of MDFT.
Of course they can be reduced, like averaged over orientations to
produce site-site radial distribution functions $g(r)$. The ideal
functional (eq.~\ref{eq:fid}) is known exactly. The external functional
(eq.~\ref{eq:fext}) reflects the direct solute-solvent interaction
energy $\phi\left(\boldsymbol{r},\omega\right)$, which is typically
given by classical force fields like SPC/E water for the solvent~\cite{berendsen_missing_1987}
and OPLS for the solute~\cite{jorgensen_opls_1988,jorgensen_development_1996}.
As usual in liquid state theories, the excess contribution (eq.~\ref{eq:fexc})
is not tractable numerically. It is often approximated by a second
order expansion around a bulk solvent density $\rho_{b}$. $\rho_{b}$
can be any density, even if it is often the density of the liquid
phase, $\rho_{\textrm{L}}$. Higher orders are gently put into a so-called
bridge functional, $\mathcal{F}_{\mathrm{b}}$. 
\begin{eqnarray}
\mathcal{F}_{\mathrm{id}} & = & \mathrm{k_{B}}T\int\mathrm{d}\boldsymbol{r}\mathrm{d}\omega\rho\left(\boldsymbol{r},\omega\right)\log\left(\frac{\rho\left(\boldsymbol{r},\omega\right)}{\rho_{b}}\right)-\rho\left(\boldsymbol{r},\omega\right)+\rho_{b},\label{eq:fid}\\
\mathcal{F}_{\mathrm{ext}} & = & \int\mathrm{d}\boldsymbol{r}\mathrm{d}\omega\rho\left(\boldsymbol{r},\omega\right)\phi\left(\boldsymbol{r},\omega\right),\label{eq:fext}\\
\mathcal{F}_{\mathrm{exc}} & = & \mathcal{F}_{\mathrm{hnc}}+\mathcal{F}_{\mathrm{b}},\\
 & = & -\frac{\mathrm{k_{B}}T}{2}\int\mathrm{d}\boldsymbol{r}\mathrm{d}\omega\Delta\rho\left(\boldsymbol{r},\omega\right)\int\mathrm{d}\boldsymbol{r}^{\prime}\mathrm{d}\omega^{\prime}c\left(\boldsymbol{r}-\boldsymbol{r}^{\prime},\omega,\omega^{\prime}\right)\Delta\rho\left(\boldsymbol{r}^{\prime},\omega^{\prime}\right)+\mathcal{F}_{\mathrm{b}},\label{eq:fexc}
\end{eqnarray}
where $\mathrm{k_{B}}$ is the Boltzmann constant, $T$ is the temperature,
and $\Delta\rho\left(\boldsymbol{r},\omega\right)\equiv\rho\left(\boldsymbol{r},\omega\right)-\rho_{b}$
is the excess density. $c$ is the direct correlation function of
the bulk solvent at a given temperature and pressure. In the following
applications, $\rho_{b}=\rho_{\textrm{L}}$ of SPC/E water at 298.15~K
and 1~atm), calculated exactly by Belloni~\cite{belloni-to-come}.
$c$ is an input of MDFT. Even in the hypernetted-chain approximation,
also called homogeneous reference fluid approximation~\cite{ramirez02}
or equivalently the Chandler-McCoy-Singer theory~\cite{chandler_density_1986,chandler_density_1986-1},
that is without a bridge functional, the calculation of the excess
term is a numerical challenge because of the spatial and angular convolutions.
We recently reported a solution that relies on projections onto a
basis of generalized spherical harmonics and using the optimal frame~\cite{dingHNC2017}.
\textit{In what follows, we build a new bridge functional $\mathcal{F}_{\mathrm{b}}$
to go beyond the HNC approximation with unpreceded accuracy and yet
numerically simple.}\textit{\emph{ The first paragraphs are dedicated
to building a bridge functional that is thermodynamically consistent.
Then, we benchmark it on the important case of water, (i) by checking
that indeed the important macroscopic thermodynamics are recovered,
and (ii) on the hydration structure and hydration free energy of hydrophobic
compounds.}}

Various bridge functionals have been proposed in the past. In particular,
bridge functionals built from the fundamental measure theory (FMT)
for hard sphere fluids have been shown to significantly improve the
predicted solvation free-energies~\cite{yu_density_2002,yu_structures_2002,zhao-wu11,zhao-wu11-correction,levesque_scalar_2012,sergiievskyi_pressure_2015}.
The criteria that make it challenging to build a bridge functional
for molecular fluids are : (i) MDFT should predict correct solvation
free energies and microscopic structures, and (ii) it should be thermodynamically
consistent for nano- macrosized solutes. FMT based bridge functionals
do not satisfy this last criteria, as shown by Sergiievskyi et al.~\cite{sergiievskyi_pressure_2015}.
As of today, despite our recent progress~\cite{jeanmairet_molecular_2013-1},
no bridge functional fulfill completely those two criteria.

\begin{figure}
\begin{centering}
\includegraphics[width=8.25cm]{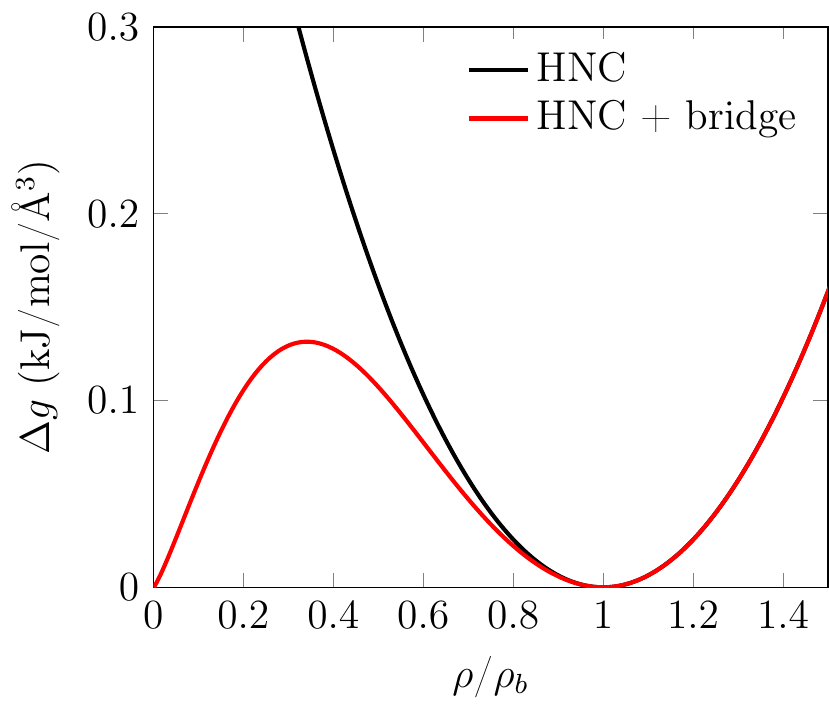}
\par\end{centering}
\caption{Predicted difference between the grand potential of homogeneous water
at a given density $\rho$ and of the grand potential of the bulk
solvent at density $\rho_{b}$. For water, $\rho_{b}$ would be $\rho_{\textrm{L}}=0.033$
water molecule per \AA$^{3}$ ($\equiv1$~kg/L at 298.15~K and
0.033~water molecule per \AA$^{3}$. In black: results from MDFT
in the HNC approximation (without bridge). In red, results from MDFT
with the coarse grained bridge functional proposed in this work. The
narrow left local minimum of the red curve is found at $\rho_{\textrm{G}}\approx5.10^{-3}\rho_{\textrm{L}}\approx0$,
as expected for water. \label{fig:fonctionelle}}
\end{figure}

To build our functional, we start from the HNC functional. In this
approximation, there is no gas phase: as shown in black in figure
\ref{fig:fonctionelle}, the free energy per volume unit fluid increases
quadraticaly with the excess fluid density. Consequently, the pressure
in the fluid is significantly overestimated \cite{rickayzen_integral_1984,jeanmairet_molecular_2015}.
It is typically 10~000~atm for water at room temperature. Said differently,
the work required to create a cavity in the fluid is overestimated,
while in real water, for instance, the liquid and gas phases are almost
at coexistence at room temperature. Recently, we proposed \textit{a-posteriori}
pressure corrections to solvation free energies predicted from the
HNC approximation. They improve significantly the correlation to reference
simulations, but are insufficient since they do not improve the solvent
density profile around the solute: They remove the overestimated work
needed to create the cavity in which the solute lies. This is not
satisfying enough and the first axis to our strategy is to make our
functional compatible with the liquid-gas coexistence. This comes
back to having no difference in volumic free energy between the gas
and liquid phase, as seen in the red curve of figure \ref{fig:fonctionelle}. 

What is the physical quantity that should drive our bridge functional
? We borrow ideas at the root of the quasi-exact classical density
functional theory for hard spheres : a weighted-density should be
able to connect the local, molecular scale, and the longer scales.
The successes of these DFTs of hard sphere fluids came with the introduction
by Nordholm, Johnson and Freasier \cite{nordholm1980} of coarse-grained,
weighted densities, and the seminal papers by Tarazona and Rosenfeld~\cite{tarazona_free-energy_1985,rosenfeld_free-energy_1989,tarazonaHSbook2008}
introducing hard-sphere measures (radius, surface, volume). For molecular
fluids, we go back to the simplest weighted density 
\begin{equation}
\bar{\rho}(\boldsymbol{r})=\int\mathrm{d}\boldsymbol{r}^{\prime}\rho(\boldsymbol{r}^{\prime})K\left(\boldsymbol{r}-\boldsymbol{r}^{\prime}\right),\label{eq:convolution}
\end{equation}
where the local number density is first averaged over all orientations,
$\rho(\mathbf{\boldsymbol{r}})=\int d\omega\rho(\mathbf{r},\omega)$,
then convoluted with a kernel $K$. The choice of such kernel is mostly
technical, and interested readers will find related information in
the supplementary material. We end up using a Gaussian kernel with
a width of 0.935~\AA.

For the expression of the bridge functional, one is already constrained
by (\emph{i}) the liquid-gas coexistence and (\emph{ii}) the connection
to HNC at the homogeneous liquid density. The first condition imposes
a double-well for the bulk free-energy as function of $\rho$, with
minima at $\rho_{\textrm{G}}\simeq0$ and $\rho_{\textrm{L}}=\rho_{b}$.
In real water, $\rho_{\textrm{G}}$ would correspond to a mass density
of $3\cdot10^{-3}$~kg/L and $\rho_{\textrm{L}}$ to 1~kg/L. As
in many field theories, the height of the barrier is linked to the
surface tension $\gamma$ between the two phases. There is nevertheless
no analytic relation between the two for our version of classical
DFT, as far as we know. Our bridge was thus constructed in two parts.
The first one, of order 3 in $\Delta\rho$, was chosen to impose $\mathcal{F}[\rho_{\textrm{G}}]\approx\mathcal{F}\left[0\right]=\mathcal{F}[\rho_{L}]$
and so to make liquid-gas coexistence possible. The second part, of
order 4 in $\Delta\rho$, should enable allow variations of the height
of the double-well without modifying the shape of the wells themselves.
Thus we chose: 
\begin{equation}
F_{\mathrm{b}}[\bar{\rho}(\boldsymbol{r})]=A\int\Delta\bar{\rho}(\boldsymbol{r})^{3}d\boldsymbol{r}+B\int\bar{\rho}(\boldsymbol{r})^{2}\Delta\bar{\rho}(\boldsymbol{r})^{4}d\boldsymbol{r}\label{eq:fbridge_2}
\end{equation}
where two parameters $A$ and $B$ are enough to sustain all criteria.
The value $A=8\pi^{2}k_{B}T\rho_{\textrm{L}}^{-2}(1-\frac{1}{2}\rho_{\textrm{L}}\hat{c}_{00;0}^{00}\left(\boldsymbol{k}=0\right))$
is imposed by the coexistence condition, where $\hat{c}_{00;0}^{00}\left(0\right)$
is the totally spherically symmetric, mean contribution of $c$. $B$
is determined using brute force try and error to get the correct surface
tension of the solvent, $\gamma=63.6$~mJ/m$^{2}$ for SPC/E water~\cite{vega_surface_2007},
leading to $B=2.37.10^{-15}k_{B}T\rho_{\textrm{L}}^{-5}$. The determination
of $A$ and $B$ is discussed in great details in the supplementary
material. With these we do have : the liquid-gas coexistence and incidentally
correct pressure and correct surface tension. The resulting free energy
curve is shown in red in figure \ref{fig:fonctionelle}. Bridge functionals
built from FMT as in~\cite{levesque_scalar_2012,zhao-wu11,zhao-wu11-correction}
or from three-body weighted-densities as introduced in \cite{jeanmairet_molecular_2013}
lead to a one-parameter adjustment (the hard sphere radius or cosine
prefactor) that could not fill all our requirements. The functional
form in equation~\ref{eq:fbridge_2} is furthermore significantly
simpler and more efficient numerically. At this point, we have a functional
that is built and tuned to be consistent from the point of view of
macroscopic thermodynamics. We now want to benchmark it on classical
test-cases without any more parameter.

\begin{figure}
\begin{centering}
\includegraphics[width=8.25cm]{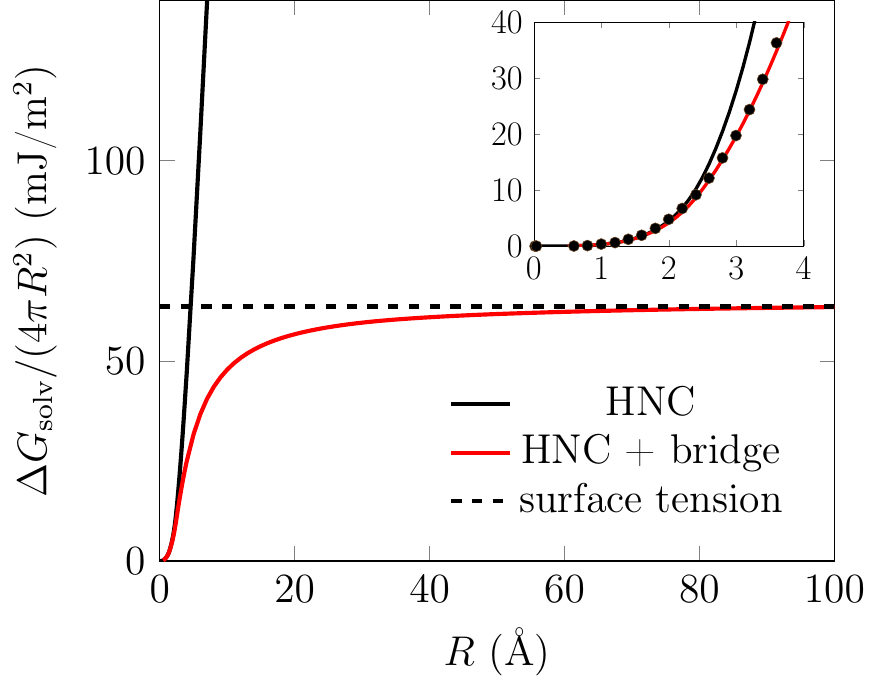}
\par\end{centering}
\caption{Free energy of hydration of a hard sphere divided by its surface,
as a function of its radius. In the insert, we focus on small radii
for which reference simulations by Hummer et al.~\cite{hummer_information_1996}
are available (black circles). The reference surface tension of SPC/E
water, 63.6~mJ.m$^{-2}$, was calculated by means of all-atom simulations
by Vega and Miguel~\cite{vega_surface_2007}. \label{fig:surface_tension} }
\end{figure}

The solvation free energy per surface unit of a simple spherical hard
sphere of various radii in SPC/E water is shown in figure \ref{fig:surface_tension}.
The reference all atom simulations by Hummer et al. \cite{hummer_information_1996}
are shown as black circles. Due to the computational demand of all
atom molecular dynamics (or Monte Carlo) simulations, only small spheres
of radius less than 4~\AA~ have been considered by the authors.
For small radii, the solvation free energy should be proportional
to the volume of the sphere~\cite{chandler_lectures_2011}, which
is the case for both reference simulations and MDFT. MDFT results
without and with the bridge functional are shown in black and red
lines, respectively. The bridge functional brings a quantitative agreement
with the reference simulations. For spheres whose diameter grows to
more than 3 to 4~\AA, that is when the interface between vacuum
and water becomes flat, the solvation free energy is driven by a surface
term. Indeed, due to the almost-coexistence, the pressure is almost
zero in the system and cancels out the volume contribution. Consequently,
the solvation free energy per surface unit of an infinitely large
sphere, that is of a hard wall, tends toward the surface tension of
the solvent. As described in the introduction, the inconsistent thermodynamics
of the HNC approximation leads to a wrong trend, driven by the volume
of the sphere. On the contrary, results corrected by the bridge functional
show the good trend and the good surface tension of the solvent, that
is, the correct limit at infinite radius $R$ in figure~\ref{fig:surface_tension}.
Fujita and Yamamoto assessed recently the accuracy of 3D-RISM for
nano-sized solutes in water~\cite{Fujita2017} and found significantly
larger discrepencies than what is showed in figure~\ref{fig:surface_tension}.
This is not surprising since 3D-RISM does not include any information
about the gas phase, as shown by Sergiievskyi et al.~\cite{sergiievskyi_solvation_2015}.

We also calculated the solvation free energy of various realistic
molecules of spherical symmetry: methane, neon, argon, krypton and
xeon. Results are shown in table \ref{tab:deltag}. Without the new
bridge functional, the solvation free energy is systematically overestimated
by large errors. With the new bridge functional, the error gets down
to within few kJ/mol of the experimental values. Let us recall here
that the parameterization of the functional does not imply any fitting
on the solvation free energies of molecules, contrarily to other bridge
functionals~\cite{yu_structures_2002,zhao-wu11,zhao-wu11-correction,levesque_solvation_2012,Sheng2016}.
Also, once again, no pressure correction \cite{sergiievskyi_pressure_2015,sergiievskyi_solvation_2015}
is needed within this new theoretical framework since the pressure
of the fluid is correct.

\begin{figure}
\begin{centering}
\includegraphics[width=8.25cm]{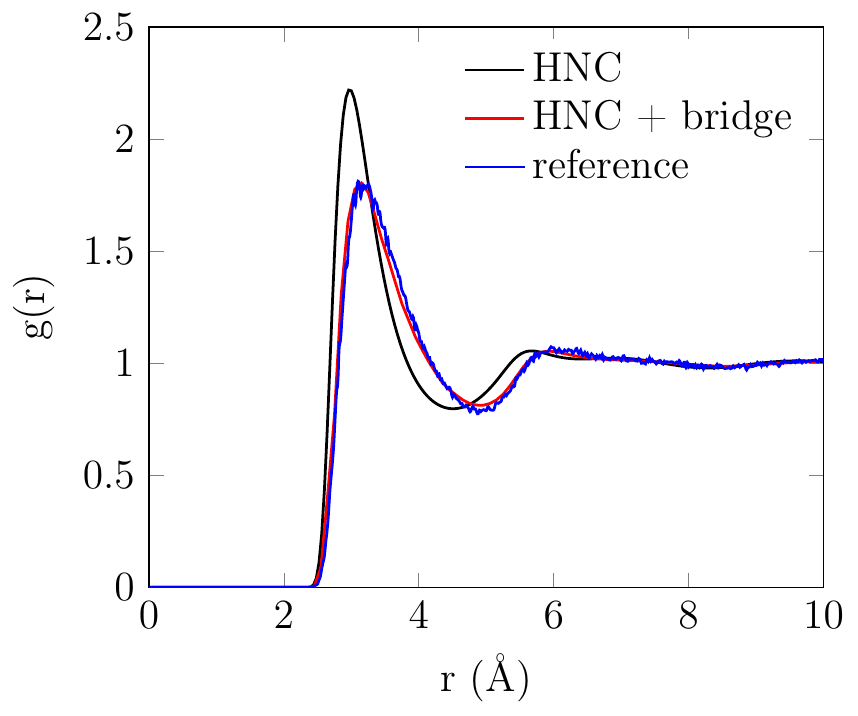}
\par\end{centering}
\caption{Radial distribution function (rdf) of neon. Those calculated by MDFT-HNC
are shown in black and with the new coarse-grained bridge functional
in red. The reference calculation shown in blue is done by the authors,
as described in supplementary information.\label{fig:g_of_r} }
\end{figure}

\begin{figure}
\begin{centering}
\includegraphics[width=8.25cm]{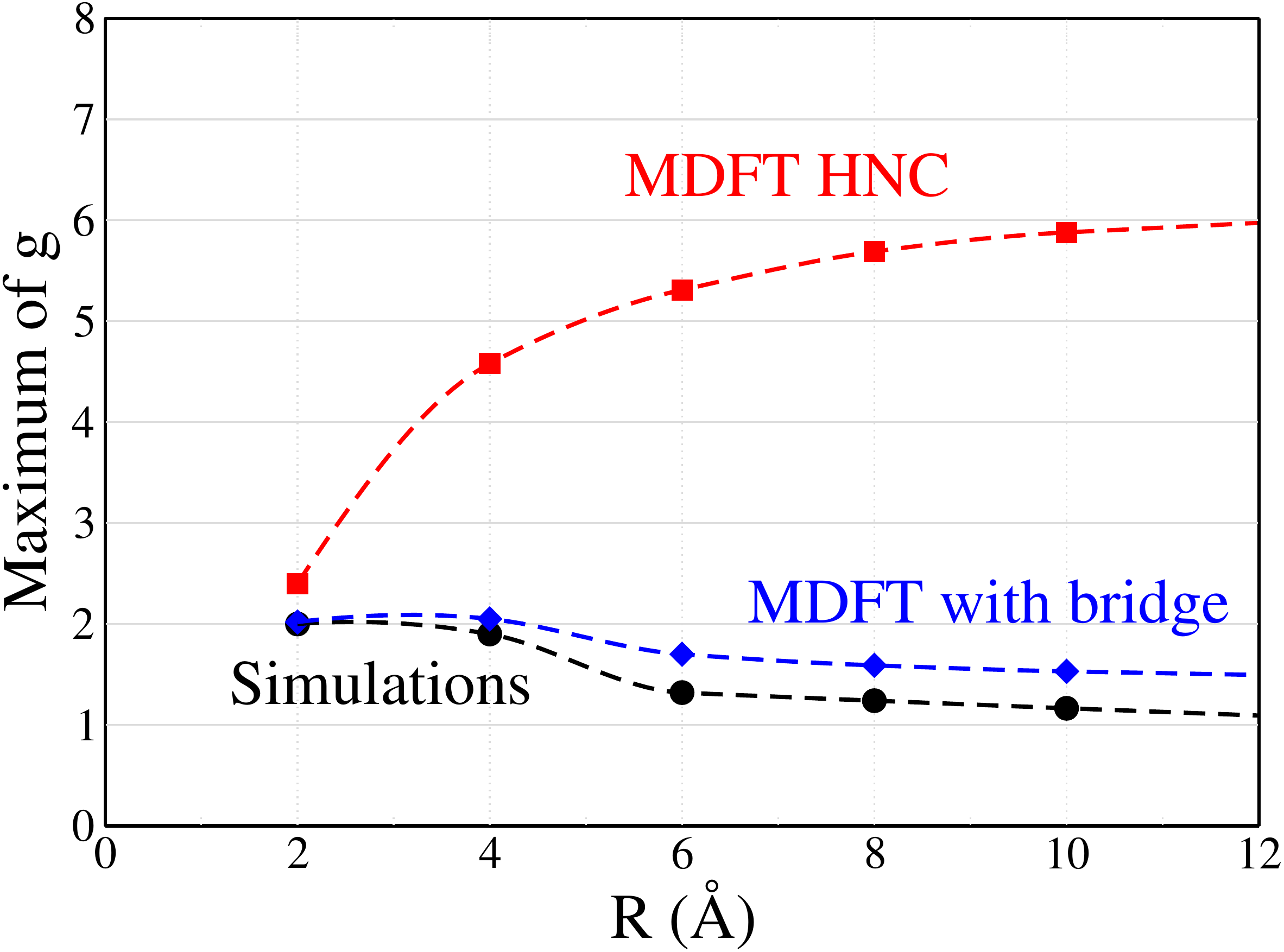}
\par\end{centering}
\caption{Height of the first peak of the radial distribution function ($g$)
between the oxygen of SPC/E water and a hard spheres of radius 2,
4, 6, 8 or 10~\AA. The reference calculations are exact Monte Carlo
simulations by Huang and Chandler~\cite{huang_hydrophobic_2002}.
The red and blue curves correspond to MDFT in the HNC approximation
and with the new coarse-grained bridge, respectively. \label{fig:g_of_r_HS}}
\end{figure}

For what concerns the solvation structure, the bridge functional improves
also significantly the results, as seen in figure \ref{fig:g_of_r}
for neon. Especially, the overestimation of the height of the first
peak of the radial distribution function (rdf), a usual failure within
HNC, is satisfactorily well corrected. This structuration of the water
is known to decrease in case of dewetting, that is, the height of
the first peak of the rdf must decrease for large hydrophobic solutes.
This behaviour is well reproduced by Monte Carlo simulations, like
those of Huang and Chandler \cite{huang_hydrophobic_2002}, as shown
in black circles in figure~\ref{fig:g_of_r_HS}. Within the HNC approximation,
the height of the first peak grows notably with hard sphere radius.
This is also corrected by the proposed bridge functional. To our knowledge,
all the bridge functionals in the litterature show spurious prepeaks
for large solutes, which is not the case here. More profiles are shown
in the supplementary material and lead to the same conclusion.

In the HNC approximation, MDFT overestimates the hydration free energies
by dozens of kJ/mol, even for small hydrophobic compounds, as shown
in table~\ref{tab:deltag}. Our new bridge functional also reduces
this error to only few kJ/mol, as it does for hard spheres of any
sizes, as already shown in figure~\ref{fig:surface_tension}. The
numerical cost, that is, the time to solution, is reduced by 4 orders
of magnitude between the all atom simulations and the MDFT.

\begin{table}
\begin{centering}
\begin{tabular}{|c|c|c|c|c|}
\hline 
\multirow{1}{*}{} & exp. & ref. sim. & HNC & HNC + bridge\tabularnewline
\hline 
\hline 
Methane & 8.08 & 8.9 & 27.10 & 12.25\tabularnewline
\hline 
Neon & 10.37 & 11.6 & 18.69 & 11.26\tabularnewline
\hline 
Argon & 8.31 & 8.6 & 22.25 & 10.92\tabularnewline
\hline 
Krypton & 6.97 & 7.7 & 25.41 & 10.99\tabularnewline
\hline 
Xenon & 6.06 & 7.3 & 29.72 & 11.26\tabularnewline
\hline 
\hline 
MUE & 0 & 2.57 & 23.61 & 3.88\tabularnewline
\hline 
\end{tabular}
\par\end{centering}
\begin{centering}
\par\end{centering}
\caption{Free energies of hydration in kJ/mol of methane and rare gas at 300~K
and 1~atm. Experimental values can be found in Straatsma, Berendsen
and Postma \cite{straatsma_free_1986} for the noble gas and from
Ben-Naim and Marcus~\cite{benNaimMarcus1984solvationThermodynamics}
for methane. The bridge functional proposed in this work decreases
the mean unsigned error (MUE) from 23.61~kJ/mol to 3.88~kJ/mol.
Reference molecular dynamics simulations with thermodynamics integration
have been performed by the authors and are described in the supplementary
information, with error bars estimated to 0.1~kJ/mol.\label{tab:deltag}}
\end{table}

\begin{figure}
\begin{centering}
\includegraphics[width=8.25cm]{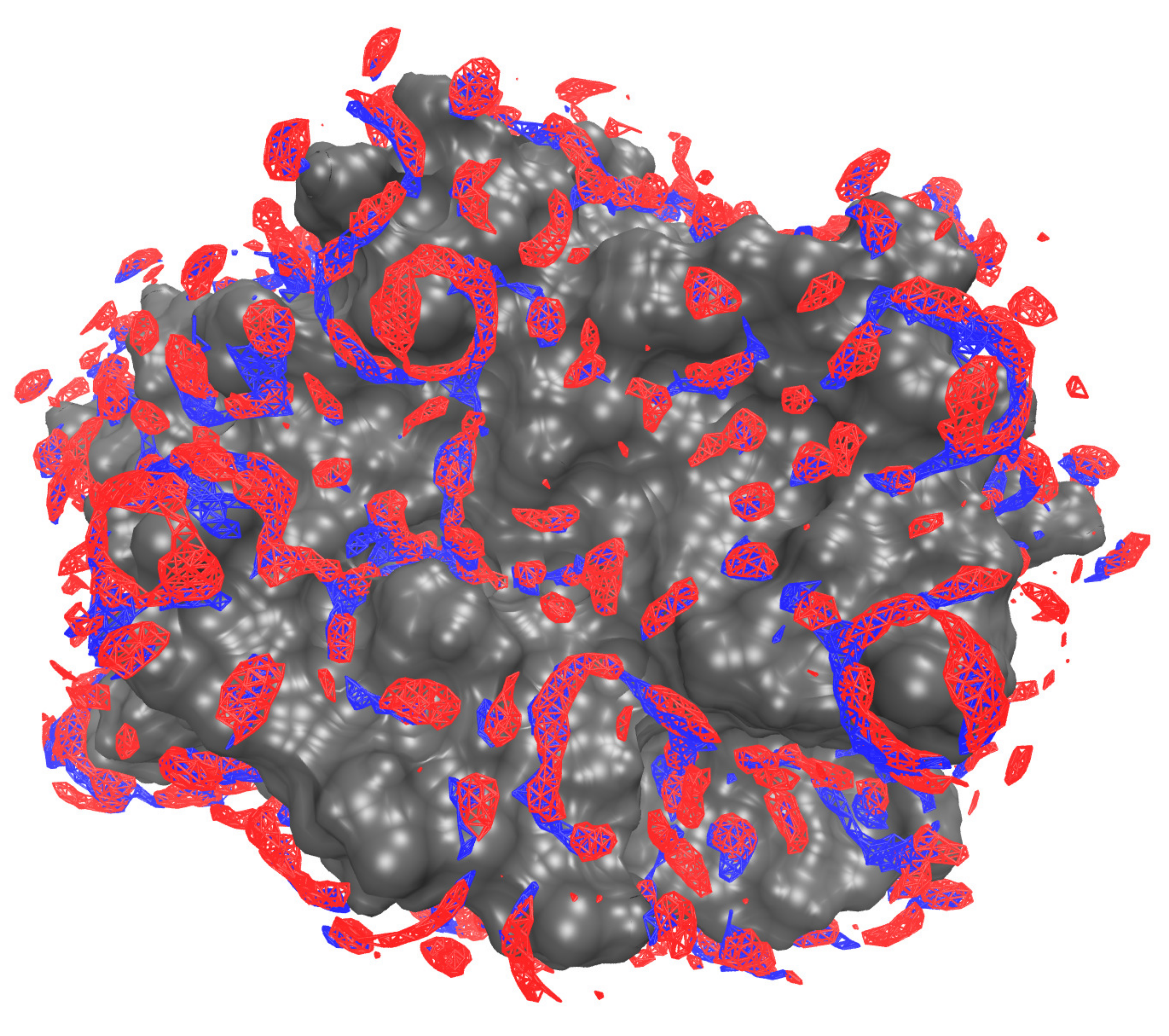}
\par\end{centering}
\caption{Hydration profile of the protein 4m7g as calculated by molecular dynamics
(blue) and MDFT (red). We plot the isosurface of high density where
the local water density is 3 times higher than it would be in the
bulk liquid, that is, where $g=3$.\label{fig:Hydration-profile-of-4m7g}}
\end{figure}

In order to illustrate the performance of this new level of theory
on a realistic molecule made of thousands of atoms, we studied the
hydration of protein 4m7g. We used both molecular dynamics (MD) and
MDFT. In figure~\ref{fig:Hydration-profile-of-4m7g}, we show the
isosurface of high densities (precisely 3 times the bulk liquid density)
as calculated by MD using a cluster for three days to gather sufficient
statistics and by MDFT with our bridge functional using a laptop with
sufficient memory in twenty minutes, in blue and red respectively.
Simulation details and the whole procedure to extract density maps
from the MD trajectories are given in supplementary materials. One
sees satisfying agreement between the two density fields. The bridge
functional presented here does improve much the results.

In this work, we have addressed the issue of predicting solvation
free energies using the molecular density functional theory (MDFT)
with a new class of bridge functionals relying on weighted densities.
We recover the liquid-gas coexistence, the correct pressure of the
liquid phase, and the liquid-gas surface tension. These ingredients
are critical for the most important case of water at ambiant pressure
and temperature. The numerical cost of the proposed coarse grained
bridge functional is negligible compared to the evaluation of the
HNC contribution, since it is made completely local in fourier space
thanks to fast Fourier transforms. Finaly and importantly, this bridge
functional improves significantly  the solvation profiles and solvation
free energies calculated by the molecular density functional theory,
at the microscopic as well as at the mesoscopic scales. This bridge
functional will be a basic ingredient for the future developments
of the molecular density functional theory. Especially, it will help
to upscale to other methods.
\begin{acknowledgments}
CG and ML thank Nicolas Chéron for discussions about the molecular
dynamics simulations of proteins. This work was supported by the Energy
oriented Centre of Excellence (EoCoE), grant agreement number 676629,
funded within the Horizon2020 framework of the European Union.
\end{acknowledgments}

\bibliographystyle{ieeetr}
\bibliography{main}

\end{document}


\title{Supplementary information for ``Bridge functional for the molecular
density functional theory with consistent pressure and surface tension
and its importance for solvation in water''}

\author{Cédric Gageat}

\affiliation{PASTEUR, Département de chimie, École normale supérieure, UPMC Univ. Paris
06, CNRS, PSL Research University, 75005 Paris, France}

\affiliation{Sorbonne Universités, UPMC Univ. Paris 06, École normale supérieure,
CNRS, PASTEUR, 75005 Paris, France}

\author{Luc Belloni}

\affiliation{LIONS, NIMBE, CEA, CNRS, Université Paris-Saclay, 91191 Gif-sur-Yvette,
France}

\author{Daniel Borgis}

\affiliation{PASTEUR, Département de chimie, École normale supérieure, UPMC Univ. Paris
06, CNRS, PSL Research University, 75005 Paris, France}

\affiliation{Sorbonne Universités, UPMC Univ. Paris 06, École normale supérieure,
CNRS, PASTEUR, 75005 Paris, France}

\affiliation{Maison de la Simulation, CEA, CNRS, Univ. Paris-Sud, UVSQ, Université
Paris-Saclay, 91191 Gif-sur-Yvette, France}

\author{Maximilien Levesque}

\affiliation{PASTEUR, Département de chimie, École normale supérieure, UPMC Univ. Paris
06, CNRS, PSL Research University, 75005 Paris, France}

\affiliation{Sorbonne Universités, UPMC Univ. Paris 06, École normale supérieure,
CNRS, PASTEUR, 75005 Paris, France}
\email{maximilien.levesque@ens.fr}

\maketitle

\section{Our functional}

The free energy molecular density functional we minimize is
\begin{equation}
\mathcal{F}=\mathcal{F}_{\textrm{id}}+\mathcal{F}_{\textrm{ext}}+\mathcal{F}_{\textrm{HNC}}+\mathcal{F}_{\textrm{b}},\label{eq:full}
\end{equation}
which is the difference between the system made of pure liquid solvent
and the system made of the solute in the solvent. In what follows,
we illustrate everything in the important case of water. In the main
manuscript, we give the following expression for the bridge functional:
\begin{equation}
F_{\mathrm{b}}[\bar{\rho}(\boldsymbol{r})]=A\int\Delta\bar{\rho}(\boldsymbol{r})^{3}d\boldsymbol{r}+B\int\bar{\rho}(\boldsymbol{r})^{2}\Delta\bar{\rho}(\boldsymbol{r})^{4}d\boldsymbol{r},\label{eq:Fb}
\end{equation}
where 
\begin{equation}
\bar{\rho}(\boldsymbol{r})\equiv\int\mathrm{d}\boldsymbol{r}^{\prime}\rho(\boldsymbol{r}^{\prime})K\left(\boldsymbol{r}-\boldsymbol{r}^{\prime}\right)\label{eq:convolution}
\end{equation}
is a weighted density, defined as the convolution between the local
solvent density $\rho\left(\boldsymbol{r}\right)$ and the convolution
kernel $K$. We note $\rho_{\textrm{G}}$ and $\rho_{\textrm{L}}$
the densities of the homogeneous gas and liquid phases. For water,
they translate to $\approx10^{-3}$~kg/L and $\approx1$~kg/L for
the gas and liquid phases, respectively. The excess (with respect
to liquid) density is defined as $\Delta\rho\left(\boldsymbol{r}\right)\equiv\rho\left(\boldsymbol{r}\right)-\rho_{\textrm{L}}$.
For what follows, we approximate the gas density to 0, that is $\rho_{\textrm{G}}\approx0$.
That's a good approximation for water for which the ratio between
vapor and liquid density is below $10^{-3}$.

\subsection{Imposing the liquid-gas coexistence}

If liquid and gas were at coexistence, which is almost the case for
water at room pressure and temperature, the difference in grand potential
of the two phases would be zero. In the very good approximation that
the gas phase has density that tends to zero, this comes to annihilate
the grand potential of the zero density limit. For what concerns the
ideal part,

\begin{equation}
\lim_{\rho\rightarrow\rho_{\textrm{G}}}\mathcal{F}_{\textrm{id}}\left[\rho\right]=k_{B}T\rho_{\textrm{L}},\label{eq:A_id}
\end{equation}
where $k_{B}$ is the Boltzmann constant and $T$ is the temperature.
There is no external contribution in the bulk fluid. For what concerns
the excess part,
\begin{equation}
\lim_{\rho\rightarrow\rho_{\textrm{G}}}\mathcal{F}_{\textrm{exc}}\left[\rho\right]=-\frac{k_{B}T}{2}\rho_{\textrm{L}}^{2}\hat{c}_{00;0}^{00}\left(0\right),\label{eq:A_exc}
\end{equation}
where $\hat{c}_{00;0}^{00}\left(0\right)$ is the totally spherically
symmetric, mean contribution of the direct, pair correlation function
of the bulk fluid. The detailed calculation of this value has been
published recently by Belloni for SPC/E water~\cite{belloni-to-come}.
This value is directly linked to the isothermal compressibility~\cite{hansen_theory_2013}.
Plugging equations~\ref{eq:A_id}, \ref{eq:A_exc} and \ref{eq:Fb}
into \ref{eq:full} and cancelling \ref{eq:full}, we have 
\begin{equation}
A=k_{B}T8\pi^{2}\frac{1}{\rho_{\textrm{L}}^{2}}\left(1-\frac{\rho_{\textrm{L}}}{2}\hat{c}_{00;0}^{00}\left(0\right)\right).
\end{equation}

\subsection{Imposing the surface tension}

\begin{figure}
\begin{centering}
\includegraphics[width=8.25cm]{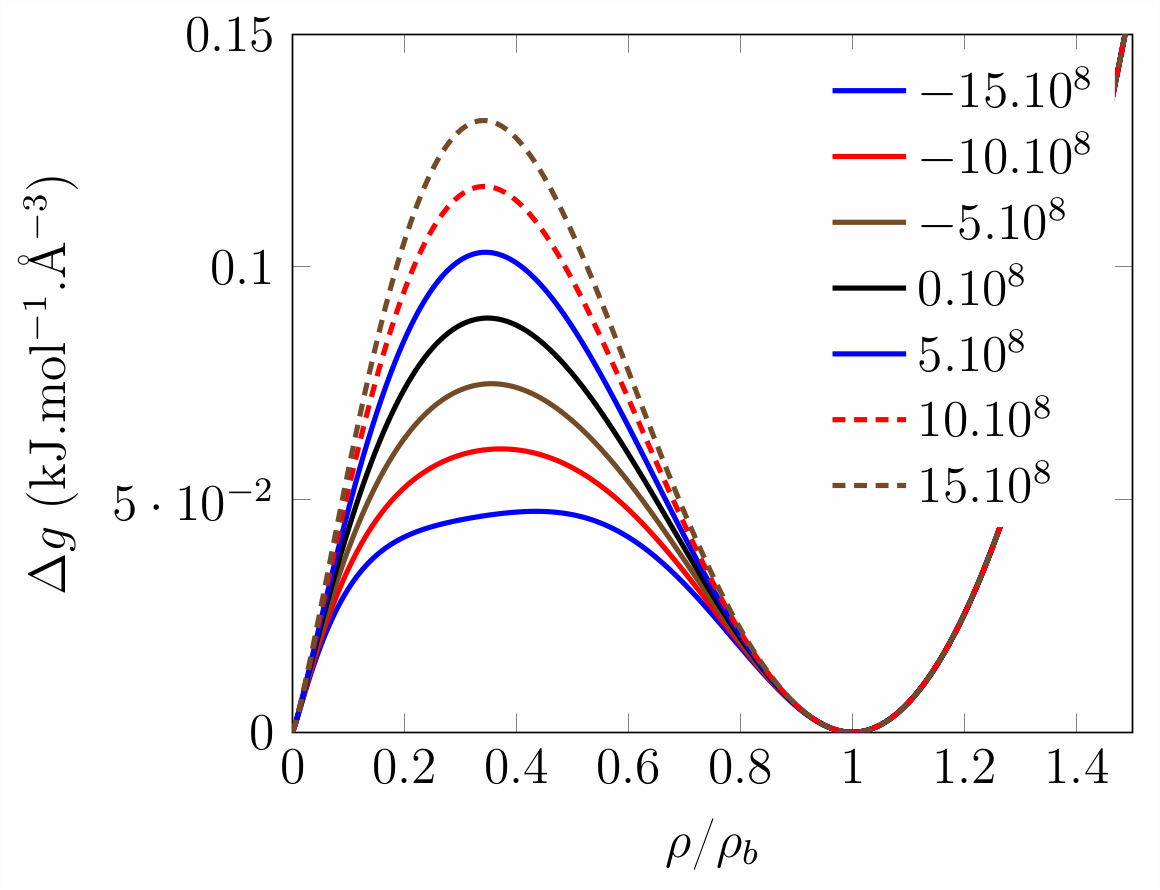}
\par\end{centering}
\caption{Influence of the parameter B of equation~\ref{eq:Fb} in the difference
in grand potential between the gas and liquid phases of SPC/E water.
The different values of B are expressed in kJ$\cdot$mol$^{-1}\cdot$\AA$^{15}$.
\label{fig:Bfig}}
\end{figure}

The parameter $B$ in equation~\ref{eq:Fb} controls the height of
the barrier between the gas and liquid phases, as illustrated in figure~\ref{fig:Bfig}.
$B$ is chosen so that the correct liquid-gas surface tension $\gamma$
is recovered. For SPC/E water, the liquid-gas surface tension $\gamma$
has been measured numerically at 63.6~mJ/m$^{2}$ by Vega and Miguel
\cite{vega_surface_2007}. We compute the surface tension by measuring
the aera-dependency of the solvation free energy of larger and larger
hard spheres of radius $R$, that is of
\begin{equation}
\gamma=\lim_{R\rightarrow\infty}\frac{\Delta G_{\textrm{hydration}}^{\textrm{hard sphere}}\left(R\right)}{4\pi R^{2}}.
\end{equation}
At this point, no more ``semi-empirical parameters'' are free in
the model. Nevertheless, in equation~\ref{eq:convolution}, we define
a weighted density where a convolution appears between the local density
and the kernel. We tried heavyside and gaussian kernels. Heavysides
led to poor results that are not shown therein. Even if at this point
no more parameter is free, one hidden numerical parameter must be
set : the gaussian distribution's width. For each possible value of
parameter $B$, only one gaussian width gives the right surface tension.
After brute-force try and error of all possible couples of $B$ and
gaussian width, we end-up with a gaussian width $B=2.37.10^{-15}k_{B}T\rho_{\textrm{L}}^{-5}$
and a gaussian width of 0.935~\AA which also optimizes the solvation
free energies, as shown in the main text.

\section{Reference simulations and force field parameters}

In order to have reference results to compare with MDFT, we run molecular
dynamics simulations. We use Gromacs~\cite{berendsen_gromacs:_1995}.
Solutes are solvated in SPC/E water in a simulation box of width 20~\AA.
We always follow the same, standard process : we start by minimizing
the internal energy of the system. Then, the simulation runs for 1~ns
at constant volume and constant temperature of $298.15$~K, followed
by a constant pressure simulation at 1~atm for about 2~ns. Finally,
we start the production run of $100$~ns from which we extract the
radial distribution function(s). For computing reference solvation
free energies, we use the weighted histogram approximation method
(WHAM) implemented in Gromacs~\cite{KumarRosenberg_WHAM_1992}. It
needs 21 steps for a coupling parameter varying from 0 to 1 with a
step of 0.05. For this free energy runs, each frame runs for 1~ns.
All input files are publicly available on github~\cite{cedricGromacsInputs}.
The Lennard-Jones parameters used in these simulations are given in
table~\ref{tab:Lennard-Jones-parameters-used}.

\begin{table}
\begin{centering}
\begin{tabular}{|c|c|c|}
\hline 
 & $\sigma$ (\AA) & $\epsilon$ (kJ/mol)\tabularnewline
\hline 
\hline 
methane\cite{asthagiri_role_2008} & 3.730 & 1.230\tabularnewline
\hline 
neopentane\cite{sarma2011hydrophobic} & 6.150 & 3.496\tabularnewline
\hline 
neon\cite{guillot_computer_1993} & 3.035 & 0.154\tabularnewline
\hline 
argon\cite{guillot_computer_1993} & 3.415 & 1.039\tabularnewline
\hline 
krypton\cite{guillot_computer_1993} & 3.675 & 1.405\tabularnewline
\hline 
xenon\cite{guillot_computer_1993} & 3.975 & 1.785\tabularnewline
\hline 
\end{tabular}
\par\end{centering}
\caption{Lennard-Jones parameters used in the molecular dynamics simulations.
These parameters are mixed to SPC/E parameters by usual Lorentz-Berthelot
mixing rules.\label{tab:Lennard-Jones-parameters-used}}

\end{table}

For the special case of the large protein 4m7g presented in figure~5
of the main manuscript, we use the GROMOS97 force field. Also, we
froze the coordinatess of each site of the protein. The density maps
are extracted from snapshots every 10~ps of the 100~ns simulation.
This last calculation took 2 days on a high-end workstation with 32
cpu-cores and 2 middle-end GPUs.
\begin{acknowledgments}
CG and ML thank Nicolas Chéron for discussions about the molecular
dynamics simulations. This work was supported by the Energy oriented
Centre of Excellence (EoCoE), grant agreement number 676629, funded
within the Horizon2020 framework of the European Union.
\end{acknowledgments}

\bibliographystyle{ieeetr}
\bibliography{main}